\input{aipcheck}
\documentclass[
    ,final            
  ]
  {aipproc}

\layoutstyle{8x11single}

\usepackage[cp866]{inputenc}
\usepackage{amsmath}
\usepackage{times}
\usepackage{amsbsy}
\usepackage[dvips]{epsfig}
\usepackage{graphicx}
\usepackage{epsfig}
\newcommand{\vk}{\vec{k}}

\newcommand{\vks}{\vec{k}^{\;2}}
\newcommand{\q}{\vec{q}}
\newcommand{\qs}{\vec{q}^{\;2}}

\def\pnot{\mbox{${\not{\hbox{\kern-3.0pt$p$}}}$}}
\def\qnot{\mbox{${\not{\hbox{\kern-2.0pt$q$}}}$}}
\def\enot{\mbox{${\not{\hbox{\kern-2.0pt$e$}}}$}}
\def\knot{\mbox{${\not{\hbox{\kern-2.0pt$k$}}}$}}

\newcommand{\be}{\begin{equation}}
\newcommand{\ee}{\end{equation}}
\newcommand{\ben}{\begin{equation*}}
\newcommand{\een}{\end{equation*}}
\newcommand{\ar}{\begin{array}}
\newcommand{\arn}{\end{array}}

\begin{document}

\title{Discontinuities of  multi-Regge amplitudes}

\author{V.S. Fadin}{address={Budker Institute of Nuclear Physics of SD RAS
and Novosibirsk State University, 630090 Novosibirsk, Russia}}

\begin{abstract}
 In the BFKL approach,  discontinuities of multiple production  amplitudes in invariant masses of produced particles are discussed. It turns out  that  they are  in  evident  contradiction with   the BDS ansatz for $n$-gluon amplitudes in the planar $N$=4 SYM  at  $n\ge 6$. An explicit expression for the NLO discontinuity of the two-to-four amplitude in  the invariant mass of two  produced gluons is  is presented.
\end{abstract}

\maketitle

\section{Introduction}

The BFKL (Balitsky-Fadin-Kuraev-Lipatov) equation~\cite{BFKL} was derived  using the assumption of  the multi-Regge form of multiple production amplitudes. According to this assumption,  the amplitude   ${\cal A}_{{2\rightarrow 2+n}}$ of the  production of the gluons $G_1,  G_2,... G_n$ in  the  collision of particles (gluons or quarks) $A$ and $B$  in the multi-Regge kinematics (MRK), which means strong ordering of longitudinal momenta and limitation of  transverse momenta, takes the form 
\begin{equation}
{\cal A}_{{2\rightarrow 2+n}}=2s\, \Gamma_{A' A}^{R_1} 
\frac{\left(s_1/s^0_{1}\right)^{\omega(t_1)}}{t_1} \gamma_{R_1R_{2}}^{G_1}
\frac{\left(s_2/s^0_{2}\right)^{\omega(t_2)}}{t_2} \gamma_{R_2R_{3}}^{G_2}.....
\frac{\left(s_n/s^0_{n+1}\right)^{\omega(t_n)}}{t_n} \gamma_{R_nR_{n+1}}^{G_n}
\frac{\left(s_{n+1}/s_0\right)^{\omega(t_{n+1})}}{t_{n+1}}
\Gamma _{B'
B}^{R_{n+1}}~, \label{A}
\end{equation}
where  $\omega(t) = j(t)$,  $\;j(t)$ is the gluon Regge trajectory, $\Gamma_{P' P}^{R}$
and $\gamma_{RR'}^{P}$ are  scattering and production Reggeon vertices,  
\[
s=(p_A+p_B)^2, \;\; s_i=(k_{i-1}+k_{i})^2~, \; t_i=q_i^2,\;\;  i=1, ... n+1~, \;\; k_0=p_{A'}, \;\;  k_{n+1}=p_{B'}~, \;\;  k_l = p_{G_l}~, \;\; l=1, ... n~, \; \; 
\]
\begin{equation}
q_1= p_A-p_{A'},\;  \; q_{l+1} = q_{l}- k_l, \;\; \; 
 s\gg s_i \gg |t_i|~, \;\;  
\end{equation}
and $s^0_{i}$ are energy scales.  They  are not important in the leading logarithmic approximation (LLA), but must be agreed   with the Reggeon vertices in the next-to-leading approximation (NLA). 

Note that in the NLA the  simple factorized  multi-Regge form \eqref{A}  is valid only for real parts of the MRK amplitudes. Fortunately, only these parts are necessary for derivation of the BFKL equation in the NLA.  Note also  that for the derivation in the NLA,  the multi-Regge form  is supposed also for production of a couple of particles 
with limited invariant mass instead of  one of the particles $A', G_1, G_2, ... B'$.

The  assumption   of  the multi-Regge form   is extremely strong since an infinite number of production amplitudes is expressed in terms of the gluon Regge trajectory and several Reggeon vertices. It    was proved lately  in the  LLA~\cite{Balitskii:1979}  and in the NLA (see~\cite{Kozlov:2014gaa} and references therein). 

\section{Discontinuities of  multi-Regge amplitudes}
The real parts of the  amplitudes are expressed through the $s_{ij}$-channel  discontinuities of these amplitudes  ($s_{ij}=(k_i+k_j)^2$),  so that  it is possible to say that  the discontinuities are  more important  than the real parts.   Remind that the equation for the BFKL Pomeron  is derived from  consideration of the $s$-channel   discontinuities  of the elastic amplitudes. Note also that the discontinuities  are more complex than the real parts. Even the simplest of them, the  discontinuities of the  elastic amplitudes,  have not a simple factorized form. Instead, they  are given by the convolution of the   particle-particle impact factors and the Green's functions of two interacting Reggeized gluons, which are  determined by the BFKL equation.  

Evidently, the $s_{ij}$-channel  discontinuities  of multi-particle production amplitudes  are even more complex. In comparison with 
the $s$-channel   discontinuities  of the elastic amplitudes, they contain two additional components~\cite{Fadin:2006bj}:  impact factors for Reggeon-gluon transitions  and matrix elements of the gluon production operator between two-Reggeon states. These components are expressed through effective vertices describing  interaction of Reggeized gluons with ordinary  gluons and quarks. Now all of them are known  in the next-to-leading order (NLO).

\subsection{Proof of the multi-Regge form}
The most known application of the  $s_{ij}$-channel discontinuities is the  proof of the multi-Regge form of QCD amplitudes~\cite{Fadin:2006bj}. 
Compatibility of the $s$-channel  unitarity with the multi-Regge form leads to the bootstrap relations connecting discontinuities of the amplitudes with their real parts and the gluon trajectories: 
\begin{equation}
\frac{-1}{2\pi i}\left(\sum_{l=j+1}^{n+1}\Delta_{{jl}} -\sum_{l=0}^{j-1}\Delta_{{lj}} \right)\,
 =\,\frac{1}{2}\left(\omega(t_{j+1})-\omega(t_{j})\right)\Re
\;  {\cal A}_{2\rightarrow 2+n}~. 
\end{equation}
Here $\Delta_{ij}$ are   the  discontinuities of ${\cal A}_{2\rightarrow 2+n}$ in  the $s_{{ij}}$ channels, which 
must be calculated using  the
Reggeized form of amplitudes in the unitarity conditions.

It turns
out that fulfilment of an infinite set of the bootstrap relations
guarantees the multi-Regge form of scattering amplitudes. 
On the other hand, all bootstrap relations are fulfilled if several
conditions imposed on the Reggeon vertices and the trajectory
(bootstrap conditions) hold true. Fulfilment of  all these conditions is  proved now (see~\cite{Kozlov:2014gaa} and references therein)  not only in Quantum Chromodynamics (QCD), but in   Yang-Mills  theories containing  fermions and scalars in arbitrary representations of the colour group  with any Yukawa-type interaction.

There are other applications of the discontinuities, related to  the  BDS 
(Bern, Dixon, Smirnov)  ansatz~\cite{Bern:2005iz} for amplitudes with maximal helicity violation (MHV) in $N=4$ supersymmetric Yang-Mills theory with large number of colours $N_c$ (in the planar approximation). 
One of them is a simple proof of violation of this ansatz for 
$n$-gluon  amplitudes at  $n\ge 6$.   

\subsection{Violation of the BDS ansatz}

Indeed, let us consider the  $s_2$- channel  discontinuity  of the amplitude  $A_{2\rightarrow 4}$  of the process 
\[
A+B \rightarrow A'+G_1+G_2 +B'\,,\;\;\;\;  (p_{A'}+p_{G_1})^2= s_1\,, \;\; 
(p_{G_1}+p_{G_2})^2=s_2\,, \;\; (p_{G_2}+p_{B'})^2=s_3\;,
\]
\begin{equation}
p_{A}-p_{A'}=q_1\,, \;\;p_{B'}-p_{B}=q_3\,, \;\; p_{A}-p_{A'}-p_{G_1} =p_{B'}+p_{G_2}-p_{B} =q_2 \;\; 
\end{equation}
in the multi-Regge kinematics $s\gg s_i\gg -q_i^2\simeq \qs_i$ ($\q_i$ denotes transverse to the  $(p_A, p_B)$ plane  components of $q_i$).  
In the BDS ansatz dependence of this discontinuity on the energy variables $s_i$  is determined by the product of the  Regge factors $\left({s_1}\right)^{\omega(t_1)}\left({s_2}\right)^{\omega(t_2)}\left({s_3}\right)^{\omega(t_3)}$ ($t_1 \equiv q_i^2$).  But in the BFKL approach, according to \cite{Fadin:2006bj},  it contains additional factor
\begin{equation}
{\langle G_1R_1|e^{{\hat K}_m \ln\left(\frac{s_2}{|\vk_1||\vk_2|}\right)}|G_2R_3\rangle}, \label{matrix element}
\end{equation}
where $\langle G_1R_1|$ and  $|G_2R_3\rangle $ are the impact factors for Reggeon-gluon transitions, ${\hat K}_m \equiv {\hat K} -\omega(t_2)$, ${\hat K}$ is the BFKL kernel in the adjoint representation of the colour group. 
Therefore, for agreement with the BDS, the impact factors for Reggeon-gluon transitions must be proportional to the eigenfunction  of the BFKL kernel with the eigenvalue equal to the gluon trajectory, that evidently contradicts the bootstrap conditions.  Indeed, it follows from  these  conditions that such  eigenfunction is proportional to the impact factors for particle-particle transitions, not for Reggeon-gluon transitions. The last ones evidently differ from the first  already in the leading order. 

It is worthwhile to note that the  contradiction described  above appears only for $n$-gluon amplitudes with $n\geq 6$, because  the BFKL  discontinuities of amplitudes with  $n < 6$  contain, instead of \eqref{matrix element}, only  matrix elements where at least one of the impact factors describes particle-particle transition. 

In fact, the incompleteness of the BDS ansatz at $n\geq 6$ is well known. The first indications of the incompleteness were obtained in~\cite{Alday:2007he} in the strong coupling regime using  the Maldacena hypothesis~\cite{Maldacena:1997re}  about ADS/CFT duality  and  in~\cite{Drummond:2007bm} using the hypothesis of  scattering amplitude/Wilson loop correspondence.  Then  the incompleteness was shown by direct two-loop calculations in~\cite{Bern:2008ap}. 

Moreover, disagreement of the BDS ansatz with the BFKL approach is  also known~\cite{Bartels:2008ce}.  Dignity of the consideration presented  here is its simplicity,  incomparable with the rather sophisticated analysis  performed in~\cite{Bartels:2008ce}. 

\subsection{The \textbf{\textit{s$_2$}}-channel discontinuity of the {$\mathbf{\cal{A}}_{2\rightarrow 4}$} amplitude in the NLO}

Another application of the discontinuities  is to test the hypothesis  of dual conformal invariance~\cite{Drummond:2007au}, which states that  
the MHV   amplitudes are given by the products of the  BDS  amplitudes  and  the remainder functions  depending  only on the anharmonic ratios of kinematic
invariants, and the hypothesis of  scattering amplitude/Wilson loop correspondence~\cite{Drummond:2007cf}, which states  that the remainder functions are given by expectation values of Wilson loops. Both these  hypothesis are  not proved. They can be tested by comparison of the BFKL discontinuities with the discontinuities  calculated with their use \cite{Lipatov:2010qg,Fadin:2011we}. 

The BFKL  discontinuities contain the matrix element  \eqref{matrix element}. It can be  calculated  in the NLA using   the eigenvalues of the  kernel  ${\hat K}_m$  obtained  in~\cite{Fadin:2011we},  existence of the representation where the kernel is  invariant  with respect to  the M\"{o}bius transformations in the transverse momentum space  proved in~\cite{Fadin:2013hpa}  and the   Reggeon-gluon  impact factors  in this representation  found in~\cite{Fadin:2014gra}. Using these results,  we obtain~\cite{FFP} in  the NLA  for production of gluons with positive helicities  
\[
\langle G_1R_1| e^{{\hat K}_m \ln\left(\frac{s_2}{|\vk_1||\vk_2|}\right)}|G_2R_3\rangle 
=\delta(\q_1-\vk_1-\vk_2-\q_3)g^4N_c\frac{q_1^-q_3^+}{k_1^+k_2^-}
\]
\[
\times \biggl[1+ \frac{g^2N_c\Gamma(1-\epsilon)}{(4\pi)^{2+\epsilon}}\left(-\frac12 \ln^2\left(\frac{\qs_1}
  {\qs_2}\right)
 -\frac{(\vks_1)^\epsilon}{\epsilon^2}-\frac12 \ln^2\left(\frac{\qs_3}
   {\qs_2}\right)
  -\frac{(\vks_2)^\epsilon}{\epsilon^2}+4\zeta(2)\right)\biggr] 
\] 
\[
\times\Biggl[ \frac12\sum_{n=-\infty}^{+\infty}(-1)^n\int_{-\infty}^{+\infty}d\nu \, \left(e^{\omega(\nu, n) \ln\left(\frac{s_2\qs_2}{|\q_1||\q_3||\vk_1||\vk_2|}\right)}-1\right)w^{\frac{n}{2}+i\nu}(w^*)
^{-\frac{n}{2}+i\nu}
\]
\[
\int \frac{dz_1}{\pi|z_1|^2}\frac{1}{1-z_1} \left(1+\frac{g^2N_c}{16\pi^2} I(z_1)\right)
z_1^{\frac{n}{2}+i\nu}(z_1^*)
^{-\frac{n}{2}+i\nu}
\int \frac{dz_2}{\pi|z_2|^2}\frac{1}{1-z_2^*} \left(1+\frac{g^2N_c}{16\pi^2} I^*(z_2)\right)
(z^*_2)^{\frac{n}{2}-i\nu}z_2
^{-\frac{n}{2}-i\nu}
\]
\begin{equation}
+\frac{k_1^+k_2^-}{q_3^+q_1^-}\int \frac{d\vec r}{\vec r ^{\;2}(\vec q_2-\vec r)^2}
\left( q_1 -(q_1-r)\frac{\vec q_1^{\;2}}{(\vec q_1-\vec r)^2}
 \right)^- \left(- q_3 +(q_3-r)\frac{\vec q_3^{\;2}}{(\vec q_3-\vec r)^2}
  \right)^+ \Biggr]
~,\label{result}
\end{equation}
where  $a^\pm =a_x\pm ia_y$ for any two-dimensional vector $\vec a$,     $\epsilon = (D-4)/{2}$, $D$ is the space-time dimension,   $w=k_1^+q_3^+/(k_2^+q_1^+)$, 
\[
I(z)=\frac{1-z}{8}\Biggl(
\ln\biggl(\frac{|1-z|^2}{|z|^2}\biggr)\ln\biggl(\frac{|1-z|^4}{|z|^6}\biggr)
-6{\rm Li}_2(z)+6{\rm Li}_2(z^*) 
\]
\begin{equation}
-3\ln|z|^2
\ln\frac{1-z}{1-z^*} \Biggr)-\frac12\ln|1-z|^2\ln\frac{|1-z|^2}{|z|^2} -\frac38\ln^2|z|^2~,  
\end{equation}
\[
\omega (\nu , n)=\frac{g^2N_c}{8\pi^2}
\left(\frac{1}{2}\,\frac{|n|}{\nu ^2+\frac{n^2}{4}}-
\psi (1+i\nu -\frac{|n|}{2})+\psi (1-i\nu +\frac{|n|}{2})
+2\psi (1) \right)\left(1-\frac{g^2N_c}{8\pi^2}\zeta (2)\right)
\]
\[
+\left(\frac{g^2N_c}{8\pi^2}\right)^2\Biggl(\frac{1}{4}\left(\psi ^{\prime \prime}(1+i\nu +\frac{|n|}{2})+
\psi ^{\prime \prime}(1-i\nu +\frac{|n|}{2})
\frac{2i\nu \left(\psi '(1-i\nu +\frac{|n|}{2})-\psi '(1+i\nu
+\frac{|n|}{2})\right)}{\nu ^2+\frac{n^2}{4}}
\right)
\]
\begin{equation}
+3\zeta (3)+\frac{1}{4}\,\frac{|n|\,\left(\nu
^2-\frac{n^2}{4}\right)}{\left(\nu
^2+\frac{n^2}{4}\right)^3}\Biggr)~.
\end{equation}
Here $\psi (x)=(\ln \Gamma (x))'$, $\zeta (n)$ is Riemann zeta function, 
\begin{equation}
Li_2(z)=-\int_0^1\frac{dx}{x}\lvert(1-xz)~.
\end{equation}
Note, that $\omega (\nu, n)$ has the important property
\begin{equation}
\omega (0,0)=0\,
\end{equation}
in accordance with the bootstrap conditions.

Compatibility of this result with  the BDS ansatz corrected by the remainder factor calculated using the  dual conformal invariance  and   scattering amplitude/Wilson loop correspondence hypotheses  is under consideration.  

\section{Summary}
The $s_{ij}$-channel discontinuities of the MRK amplitudes are more complex  objects than the real parts of these amplitudes. They are  expressed in terms of the BFKL kernel,  the  impact factors for particle-particle and Reggeon-gluon transitions and the matrix elements of the gluon production operator between two-Reggeon states. All ingredients entering in this expressions are known now in a closed form. 

The most known application of the  discontinuities is the  proof of the multi-Regge form of the MRK amplitudes. Now this form is proved in   theories with  fermions and scalars in arbitrary representations of the colour group  with any Yukawa-type interaction. 

Knowledge of the discontinuities permits to check  the  BDS ansatz and the hypotheses about the remainder functions to this ansatz. 

The BFKL discontinuities are in  an evident contradiction with  the BDS ansatz for $2\rightarrow 2+n$ amplitudes at $n\geq 2$. 

To check the hypotheses about the remainder functions we calculated the matrix element \eqref{matrix element} in the NLA.  Comparison of the result  \eqref{result} with the results  obtained using these hypotheses is under consideration.

\begin{theacknowledgments}
The work is supported by the Ministry of Education and Science of the Russian Federation
and by the RFBR grant 13-02-01023.
The participation of V.F. to Diffraction 2014 was partially supported by
HadronPhysics3 within the Seventh Framework Programme (FP7) of EU.
\end{theacknowledgments}

\bibliographystyle{aipproc}   

\begin{thebibliography}{9}

\bibitem{BFKL}
V.~S.~Fadin, E.~A.~Kuraev and L.~N.~Lipatov,
Phys.\ Lett.\  B {\bf 60} (1975) 50; \\
E.~A.~Kuraev, L.~N.~Lipatov and V.~S.~Fadin,
Sov.\ Phys.\ JETP {\bf 44} (1976) 443,
 {\it ibid.} {\bf 45} (1977) 199; \\
I.~I.~Balitsky and L.~N.~Lipatov,
Sov.\ J.\ Nucl.\ Phys.\  {\bf 28} (1978) 822.
\bibitem{Balitskii:1979}
Ya.Ya.~Balitskii, L.N.~Lipatov and V.S.~Fadin, in {\it Materials of
IV Winter School of LNPI} (Leningrad, 1979) 109.
\bibitem{Kozlov:2014gaa}
  M.G.~Kozlov, A.V.~Reznichenko and V.S.~Fadin,
  Phys.\ Atom.\ Nucl.\  {\bf 77} (2014) 251.
\bibitem{Fadin:2006bj}
V.~S.~Fadin, R.~Fiore, M.~G.~Kozlov and A.~V.~Reznichenko,
Phys.\ Lett.\  B {\bf 639} (2006) 74.
\bibitem{Bern:2005iz}
Z.~Bern, L.J.~Dixon and V.A.~Smirnov,
Phys.\ Rev.\ D {\bf 72} (2005) 085001 [hep-th/0505205].
\bibitem{Alday:2007he}
  L.~F.~Alday and J.~Maldacena,
  JHEP {\bf 0711} (2007) 068
  [arXiv:0710.1060 [hep-th]].
\bibitem{Maldacena:1997re}
  J.~M.~Maldacena,
  Adv.\ Theor.\ Math.\ Phys.\  {\bf 2} (1998) 231
  [hep-th/9711200].
\bibitem{Drummond:2007bm}
  J.~M.~Drummond, J.~Henn, G.~P.~Korchemsky and E.~Sokatchev,
  Phys.\ Lett.\ B {\bf 662} (2008) 456
  [arXiv:0712.4138 [hep-th]].
\bibitem{Bern:2008ap}
  Z.~Bern, L.~J.~Dixon, D.~A.~Kosower, R.~Roiban, M.~Spradlin, C.~Vergu and A.~Volovich,
  Phys.\ Rev.\ D {\bf 78} (2008) 045007
  [arXiv:0803.1465 [hep-th]].
\bibitem{Bartels:2008ce}
  J.~Bartels, L.~N.~Lipatov and A.~Sabio Vera,
  Phys.\ Rev.\ D {\bf 80} (2009) 045002
  [arXiv:0802.2065 [hep-th]].
\bibitem{Drummond:2007au}
  J.~M.~Drummond, J.~Henn, G.~P.~Korchemsky and E.~Sokatchev,
  Nucl.\ Phys.\ B {\bf 826} (2010) 337
  [arXiv:0712.1223 [hep-th]].
\bibitem{Drummond:2007cf}
  J.~M.~Drummond, J.~Henn, G.~P.~Korchemsky and E.~Sokatchev,
  Nucl.\ Phys.\ B {\bf 795} (2008) 52
  [arXiv:0709.2368 [hep-th]].
\bibitem{Lipatov:2010qg}
L.N.~Lipatov and A.~Prygarin,
Phys.\ Rev.\ D {\bf 83} (2011) 045020; 
{\it ibid.} {\bf 83} (2011) 125001; 
J.~Bartels, L.N.~Lipatov and A.~Prygarin,
Phys.\ Lett.\ B {\bf 705} (2011) 507.
\bibitem{Fadin:2011we}
V.S.~Fadin and L.N.~Lipatov,
Phys.\ Lett.\ B {\bf 706} (2012) 470 [arXiv:1111.0782 [hep-th]].
\bibitem{Fadin:2013hpa}
  V.~S.~Fadin, R.~Fiore, L.~N.~Lipatov and A.~Papa,
  Nucl.\ Phys.\ B {\bf 874} (2013) 230.
\bibitem{Fadin:2014gra}
  V.~S.~Fadin and R.~Fiore,
  arXiv:1402.5260 [hep-th].
\bibitem{FFP}
  V.~S.~Fadin and  R.~Fiore, to be published.  
\end{thebibliography}

\end{document}